\documentclass{article}
\usepackage[utf8]{inputenc}
\usepackage{gensymb}
\usepackage{graphicx}
\usepackage{dblfloatfix}
\usepackage{authblk}

\usepackage{geometry}
\graphicspath{{jpegfigures/}}
\title{FeTe$_{0.55}$Se$_{0.45}$ van der Waals Tunneling Devices}
\date{}
\newcommand{\FTS}{FeTe$_{0.55}$Se$_{0.45}$}
\author[1]{\thanks{Equal contribution}Ayelet Zalic} 
\author[1]{$^{\ast}$Shahar Simon}
\author[2]{Sergei Remennik}
\author[2]{Atzmon Vakahi}
\author[3]{Genda D. Gu}
\author[1,2]{Hadar Steinberg}
\affil[1]{\textit{The Racah Institute of Physics, The Hebrew University of Jerusalem, Jerusalem 91904, Israel}}
\affil[2]{\textit{The Center for Nanoscience and Nanotechnology, The Hebrew University of Jerusalem, Jerusalem, 91904, Israel}}
\affil[3]{\textit{Brookhaven National Laboratory, Condensed Matter Physics and Materials Science Department,
Upton, NY 11973, USA}}

\begin{document}

\maketitle

\begin{abstract}
    We report on fabrication of devices integrating \FTS\ with other van-der-Waals materials, measuring transport properties as well as tunneling spectra at variable magnetic fields and temperatures down to 35 mK. Transport measurements are reliable and repeatable, revealing temperature and magnetic field dependence in agreement with prior results, confirming that fabrication processing does not alter bulk properties. However, cross-section scanning transmission microscopy reveals oxidation of the surface, which may explain a lower yield of tunneling device fabrication. We nonetheless observe hard-gap planar tunneling into \FTS\ through a MoS$_2$ barrier. Notably, a minimal hard gap of 0.5 meV persists up to a magnetic field of 9 T in the $ab$ plane and 3 T out of plane. This may be the result of very small junction dimensions, or a quantum-limit minimal energy spacing between vortex bound states. We also observed defect assisted tunneling, exhibiting bias-symmetric resonant states which may arise due to resonant Andreev processes.
\end{abstract}

\section*{Introduction}

    Superconductors of the Fe(Te,Se) family are an important research platform for unconventional superconductivity. They have electron and hole Fermi pockets at the $\Gamma$ and $M$ points which are roughly the same size, characterized by extremely small Fermi energies $E_F$. The Fermi energies are not much larger than the superconducting gap $\Delta$, resulting in a large $\Delta/E_F$ ratio. This has two important consequences: First, it makes these superconductors sensitive to interaction effects such as the BCS-BEC crossover~\cite{Lubashevsky2012,Rinott2017}. Second, the $\Delta/E_F$ ratio determines the energy separation between Caroli-de Gennes Matricon (CdGM) vortex-bound states \cite{Caroli1964}, allowing the observation of discrete CdGM states in scanning tunneling microscopy (STM) measurements~\cite{Yin2015,Fu2018,Machida2018,Kong2019,Jiang2019}. Among these, some reports observe zero energy bound states interpreted as Majorana excitations ~\cite{Wang2018,Machida2018,Kong2019}, expected in view of ARPES signatures of a topological superconducting state on the surface of \FTS~\cite{Zhang2018}. 
    
    There are several open questions related to Fe(Te,Se) superconductivity. The observed sub-gap spectra appear to vary between different studies, as well as between vortices within the same study. This variability is not yet fully understood, and neither is it clear why do some vortices within a given sample host zero energy states, while others do not. Furthermore, the symmetry of the superconducting order parameter is as yet unsettled, with evidence in favor of $S_\pm$ symmetry \cite{Hanaguri2010}. The unconventional $S_\pm$ symmetry in a topological superconductor has been suggested to give rise to chiral one-dimensional ``hinge" modes at the intersection between two faces of the superconductor \cite{zhang2018helical,Gray2019}.

    \FTS\ can be integrated into devices - for example by machining thin films into bridges~\cite{Wu2013} or by exfoliation of single crystals~\cite{Gray2019}. Refining exfoliation-based methods is of particular interest, as flakes can be assembled into a variety of sophisticated devices using the van-der-Waals (vdW) technology~\cite{Liu2016,Novoselov2016,RiccardoFrisenda2018}. When applied to proximitized graphene and to layered superconductors, stacking of vdW flakes has allowed the study of planar tunnel junctions~\cite{Bretheau2017,Dvir2018,DvirVortexState2018,Wang2018,Khestanova2018}, Josephson devices~\cite{island2016,Yabuki2016}, and graphene-superconductor hybrids \cite{Efetov2016,Sahu2016,Sahu2018}. Furthermore, in thin flakes density can be tuned by ionic gating~\cite{Costanzo2018,Liao2018}. 
    
    Here we report the integration of exfoliated single-crystal \FTS\ flakes into transport and tunneling devices. Transport devices exhibit superconducting properties which agree with bulk materials. \FTS\ is not a vdW material, and we find that exfoliated flakes oxidize rapidly even when prepared within a glove box. Despite oxidation and surface degradation which lower device yields, we are able to realize planar tunnel junctions using a van der Waals barrier and demonstrate the ability to achieve hard gap tunneling.  

\section*{Device Fabrication and Surface Quality}
\begin{figure}[h]
            \centering
            \includegraphics[width=1\textwidth]{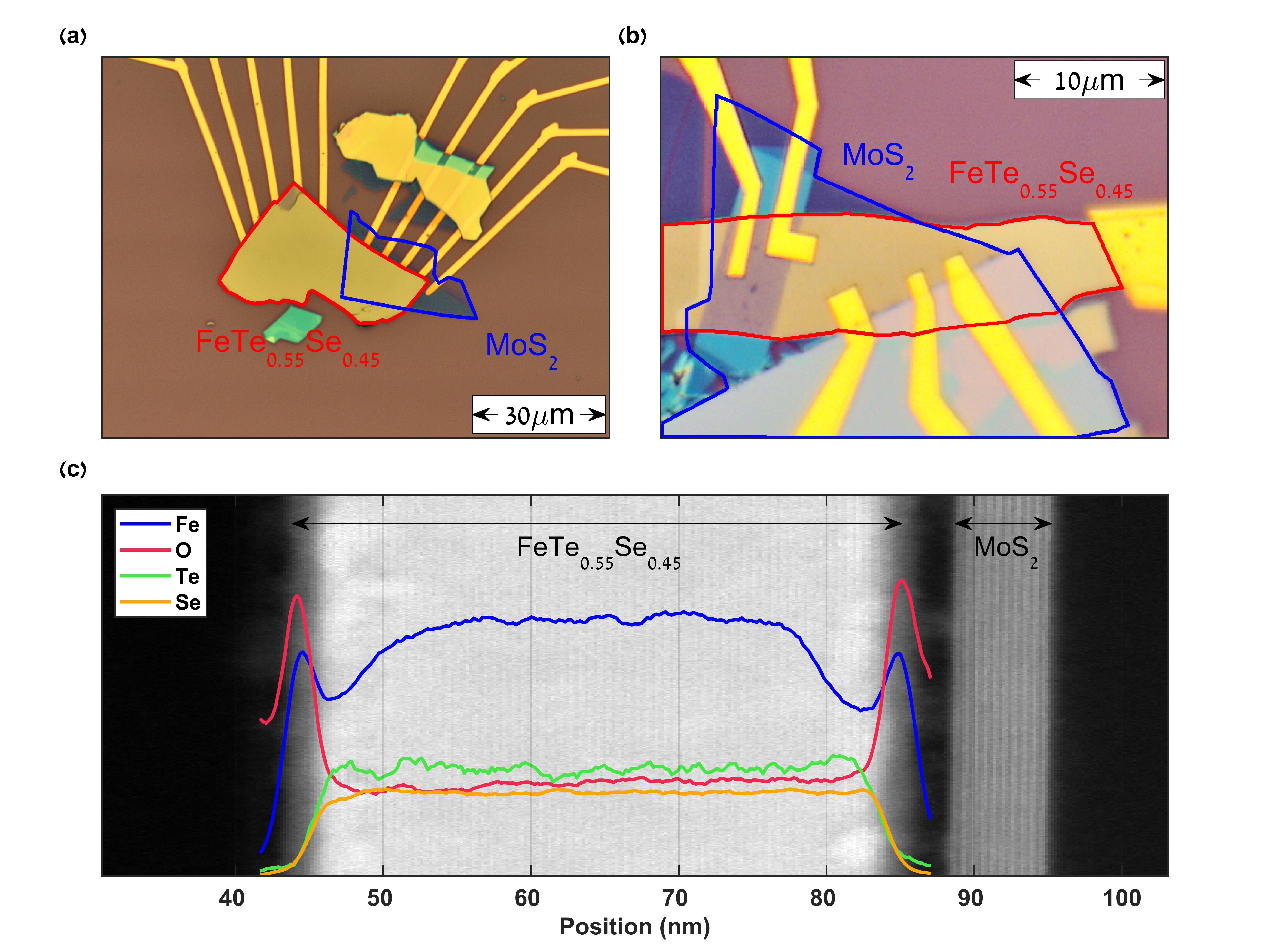}
            \caption{Tunneling devices prepared (a) on pre-patterned bottom electrodes and (b) with top electrodes using standard electron beam lithography. (c) EDS line scan, showing the relative amount of Fe, Te, Se and O in a cross-section of the device, in arbitrary units. The results are superimposed on a high angle annular dark field (HAADF) image of the same cross-section, showing layers of  \FTS\ separated by a gap from layers of the underlying MoS$_2$. }
            \label{fig:devices_TEM}
        \end{figure}

    We use large single crystals of \FTS\ grown using the self-flux method.
    All devices were prepared by exfoliation and transfer, as detailed in the methods section, in an inert N$_2$ environment inside a glove box. Data from three devices is presented, each fabricated using a different technique: 
    Device TR (transport) is fabricated by exfoliation of \FTS\ flakes on a SiO$_2$ substrate, and subsequent evaporation of electrodes.
    In Device TE (tunneling, top electrodes) \FTS\ flakes are prepared in the same way, with an MoS$_2$ tunnel barrier transferred on the top side.
    Device BE (tunneling, bottom electrodes) was fabricated differently, by placing the tunnel barrier followed by an \FTS\ flake on top of pre-patterned electrodes. 
    Images of both types of tunnel devices are shown in Figure \ref{fig:devices_TEM} (a,b) respectively. 
    The advantage of the pre-patterned electrode method is in avoiding  lithographic processing and heating of the \FTS\ and tunnel barrier, while the disadvantage is that the rough top surface of the gold does not provide continuous contact to the material. In contrast, the top-electrode fabrication method exposes the \FTS\ and barrier to polymers, solvents, and short periods of heating, but also creates a more even contact with the electrode.
    
    To better understand the behaviour of the \FTS\ at the tunnel barrier interface, we performed cross-section scanning transmission electron microscopy (STEM) and energy-dispersive spectroscopy (EDS) on a sacrificial device with a configuration identical to Device BE (details in methods section). A typical STEM measurement of \FTS\ over a barrier is shown in Figure \ref{fig:devices_TEM}(c), with an EDS line scan superimposed. In this type of device, we found that the interface between the \FTS\ and the underlying MoS$_2$ is uneven, touching only at discrete spots. Elsewhere, there is a typical spacing of around 6 nm between the layers as seen in the figure. EDS reveals variations in stoichiometry near both surfaces, with a 5 nm layer of depleted Fe followed by a  2.5 nm layer of Fe$_3$O$_4$ iron oxide\footnote{The high oxygen count within the \FTS\ is not reliable, since the oxygen $K_{\alpha}$ peak of 523 eV is within the detector resolution from a strong tellurium $M_{\xi_1}$ peak at 464 eV, which can result in an overlap between the peaks and false counts of oxygen which actually originate in tellurium.}. EDS also shows that both surfaces are coated with a layer, approximately 5 nm thick, containing silicon, carbon and oxygen (not plotted). These are assumed to originate from organic residue of the PDMS and glue used in the transfer process (see Methods section). 
        
\section*{Results}
    \subsection*{Transport}
\begin{figure}[h]
            \centering
            \includegraphics[width=1\columnwidth]{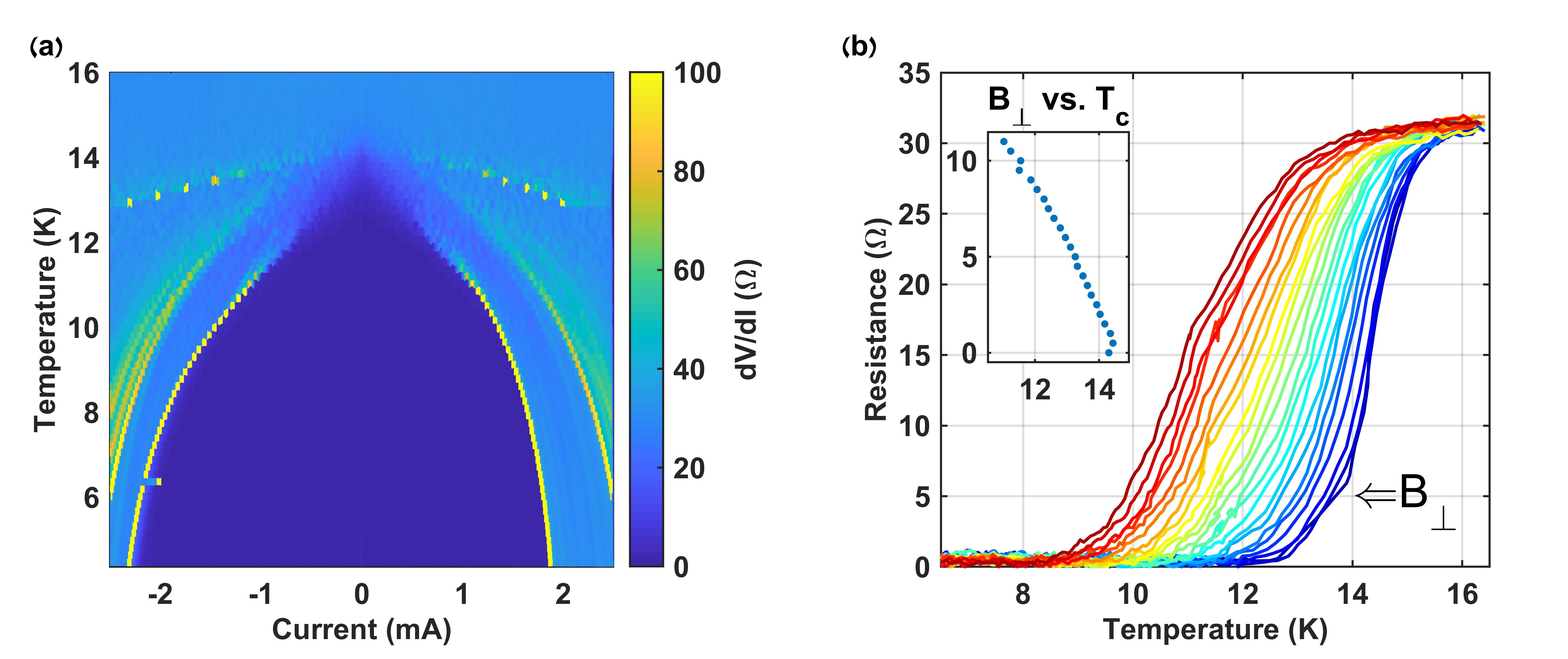}
            \caption{Transport data, measured on Device TR (100 nm thickness).
           (a) Differential resistance vs. current and temperature. Critical current decreases with temperature and disappears at around 14.5 K, in the vicinity of $T_C$. 
           (b) Resistance at 0.5 mV bias voltage, vs. $T$ at different values of $B_\perp$. As shown in the inset, $T_C$ decreases with $B_\perp$. 
           }
            \label{fig:transport}
        \end{figure}

            Transport measurements were carried out in a $^3$He cryostat with a base temperature of 300 mK.
            Figure \ref{fig:transport} shows two-terminal transport data, measured on Device TR, of thickness 100 nm. Panel (a) shows critical current measurements taken at different temperatures \footnote{Temperature was measured by two sensors located a small distance away from the sample, with a temperature gradient between them, and calibrated so that $T_C$ equals the known value for the material, 14.5 K. Electrode resistance was subtracted from the total measured resistance. }. 
            
            As expected, critical current is suppressed as $T$ increases towards $T_C$, above which the resistance is constant.
            Panel (b) shows the dependence of the critical temperature on applied  magnetic field perpendicular to the $ab$ plane ($B_\perp$), with 0.5 mV bias voltage across the sample. As expected, increasing the field lowers $T_C$. Resistance is shown relative to the minimum resistance measured at each value of $B_\perp$.
            The inset shows $T_C$ at each value of $B_\perp$, with $T_C$ defined as the point on the $T$-axis where $R$ is at half its maximum. We observed similar transport behavior in several devices with thicknesses ranging 90 -- 200 nm. These results on exfoliated flakes are consistent with transport behavior observed previously in bulk and thin film \FTS~\cite{Li2011a,Sun2015}. 
            
        \subsection*{Zero Field Tunneling Spectrum and Temperature Dependence}
\begin{figure}[h]
            \centering
            \includegraphics[width=1\columnwidth]{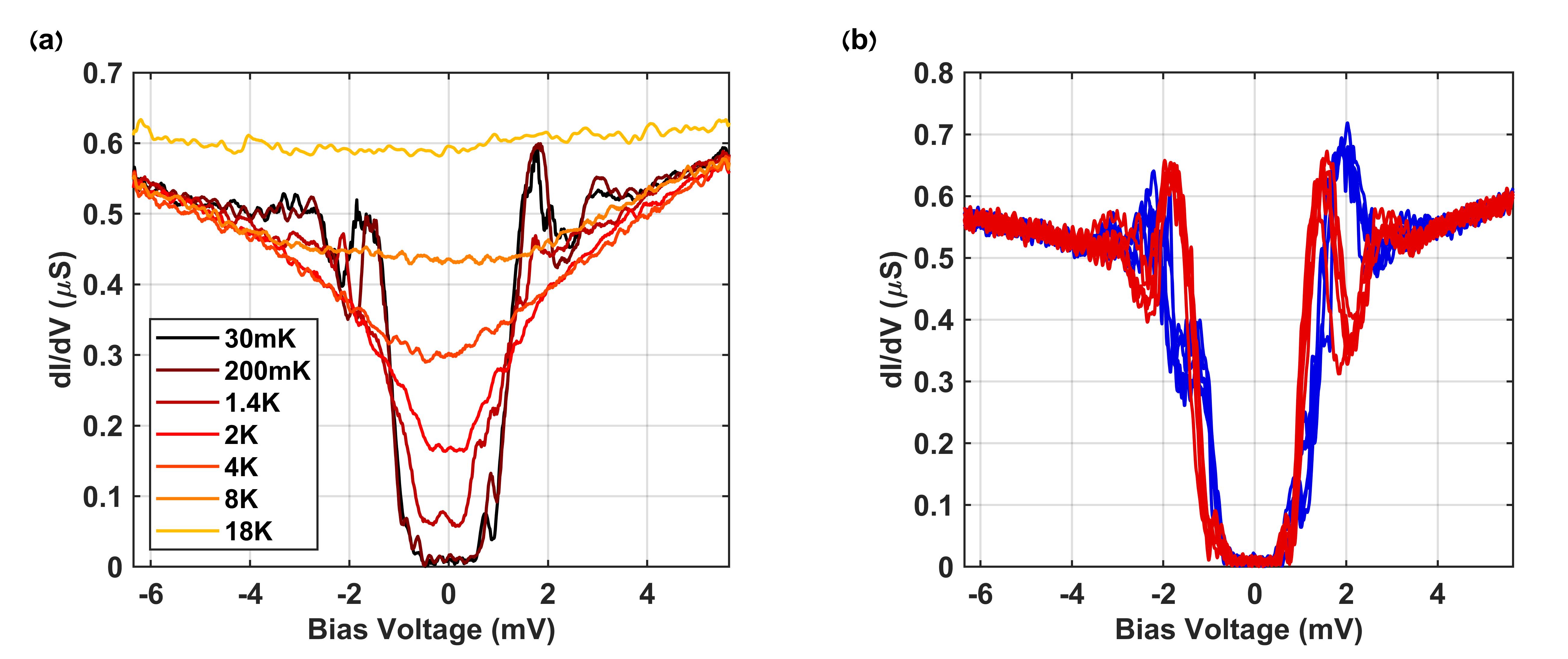}
            \caption{Tunneling spectrum of Device BE. (a) Differential conductance at different temperatures. Quasiparticle peaks flatten by 1.4 K. (b) Spectra measured consecutively at zero field and base temperature with no change of parameters, all retracing one of two modes.}
            \label{fig:tunneling}
        \end{figure}

            Tunneling measurements were carried out in a $^3$He-$^4$He dilution cryostat with a base temperature of 25 mK. Differential conductance $dI/dV$ of bottom-electrode Device BE (Figure \ref{fig:tunneling}) exhibits, at the lowest temperature, a hard superconducting gap with $R_N/R_0\approx70$. The sub-gap energy range is punctuated by in-gap states at bias voltage $V\approx\pm$0.7 meV, which we attribute to defects in the \FTS, as observed also in STM measurements~\cite{Machida2018}. 
            In addition, this junction exhibits two noteworthy phenomena: 
            First, we find that the temperature dependence (Panel (a)) shows that most of the details of the spectrum are smeared out up to T = 1.4 K, including the quasiparticle peaks, with the spectrum at 2 K showing simply a `V'-shape. This is somewhat contrasting with STM data, which shows quasiparticle peaks persisting over 5 K ~\cite{Hanaguri2010,Yin2015}.  
            Second, we find that the junction spectrum exhibits a bi-modal behavior, with the reading shifting between two different forms upon repeated measurement. Panel (b) shows several spectra, taken consecutively at the same temperature and zero magnetic field. Evidently, all of the measurements trace one of two modes of the spectrum. In one mode, the tunneling spectra reveal two sets of peaks, one at $\pm$1.7 meV, and the other at $\pm$2.7 meV, with the energy range below 1.5 meV gapped and empty of spectral weight\footnote{Note that all previously stated voltages are with respect to the center of symmetry, offset by 350 $\mu$eV from zero bias. This offset is compensated in all figures.}. The other form has three peaks at $\pm$1.5 meV, $\pm$1.9 meV, and $\pm$2.5 meV. 
            
            Comparison of these peak values to available ARPES and STM data confirms their identification as quasiparticle peaks. ARPES measurements identify four gaps: 1.7 meV and 2.5 meV for the $\alpha$' and $\beta$ hole bands at the $\Gamma$ point, 1.9 meV for the topological surface band, and 4.2 meV for the $\gamma$ electron band around the M point~\cite{Miao2012,Zhang2018}. Similar gaps have been observed in STM measurements, although the exact energies of the quasiparticle peaks vary and not all studies observe features related to all gaps \cite{Hanaguri2010,Yin2015,Fu2018,Machida2018,Kong2019}. 
            
            The quasiparticle spectrum of \FTS\ has pronounced spatial variability. Ref.~\cite{Wang2019} identifies three distinct structures: a single peak at $\pm$1.4 meV is assigned to the $\alpha'$ band, a double peak at $\pm$1.4, 2.4 meV is assigned to the $\alpha',\beta$ bands, and a triple peak at $\pm$ 1.4, 2.4, 1.9 meV is assigned to the $\alpha',\beta$ and surface state bands, respectively. In our data, we tentatively identify the double gap structure with the $\alpha',\beta$ bands, and the triple gap structure with the $\alpha',\beta$ and surface state bands. We believe the two forms of the spectrum arise when the electrode engages two different locations in the sample, although it is not clear which mechanism could generate such switching.
            
        \subsection*{Tunneling Magnetic Field Dependence}
\begin{figure}[h]
            \centering
            \includegraphics[width=1\textwidth]{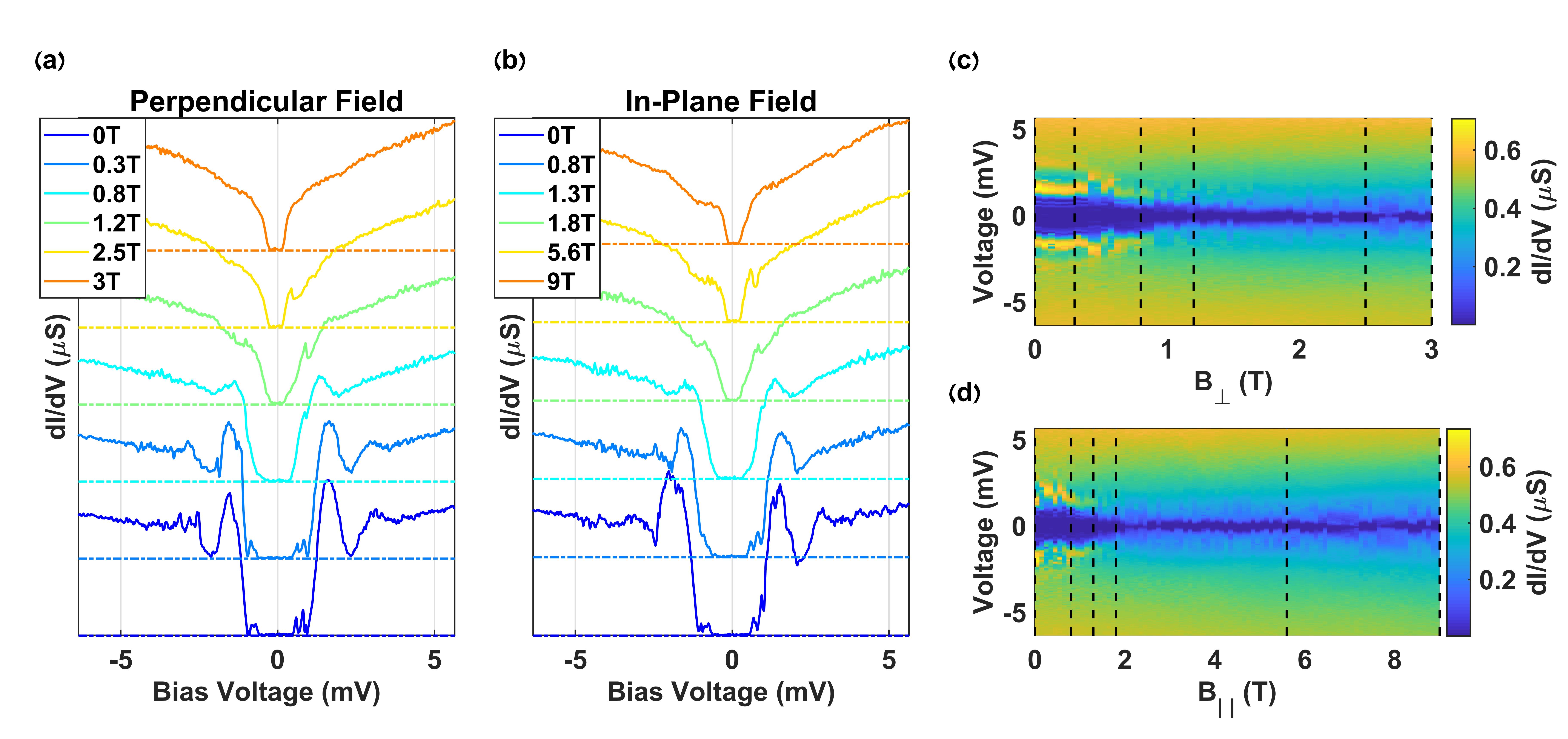}
            \caption{Response of the tunneling spectrum to $B_\perp$, $B_\parallel$. (a,b) Line scans at representative values of perpendicular and parallel fields respectively. Plots are vertically shifted for clarity.  (c,d) $dI/dV$ vs. $B_\perp$, $B_\parallel$. Vertical dashed lines in (c,d) indicate the line scans plotted in (a,b) respectively.}
            \label{fig:magnetic}
        \end{figure}

            Figure \ref{fig:magnetic} presents tunneling measurements performed on bottom electrode tunneling Device BE at different magnetic fields $B$, oriented either in the $ab$ plane of the sample ($B_\parallel$) or perpendicular to it ($B_\perp$). Measurements are realized using a 9 T--3 T vector magnet (noting that sample orientation causes the parallel field to have a small perpendicular component). Panels (a) and (b) show the spectrum at representative fields, while panels (c) and (d) show a color plot of the spectrum over the full range of magnetic fields (line scans taken for (a),(b) are marked by dashed lines). As with the zero-field spectra, one can see the alternations between the two modes here as well.
            
            As presented in panels (c) and (d), in both orientations the response of the spectrum to magnetic field is similar, with a stronger response to $B_\perp$. Below $B_\perp$ = 0.5  T / $B_\parallel$ = 1 T, the gap size does not change significantly but coherence peaks are gradually suppressed. At increasing fields the quasiparticle peaks continue to broaden and lose intensity, while spectral weight appears inside the gap. By $B_\perp$ = 1 T / $B_\parallel$ = 2 T,  the gap reaches a minimal size of around 0.5 meV for both field orientations. Upon further increase of the field, the spectrum hardly changes, aside from individual linescans showing sub-gap structure (as in the linescans for $B_\perp$ = 2.5 T / $B_\parallel$ = 5.6 T shown in panels (a) and (b)). This switching between spectral shapes may be related to the bi-modal behavior described previously. It is notable that the material retains a hard gap of 0.5 meV up to the maximal fields of $B_\perp$ = 3 T / $B_\parallel$ = 9 T.  These fields should be well over $H_{C1}$~\cite{Klein2010}, even when accounting for variability due to device geometry~\cite{DvirVortexState2018}. We believe that this field dependent behavior may arise from vortices entering the junction, with spectral weight within the gap reflecting either depairing in the vicinity of vortices due to screening currents, sub-gap states within the vortices, or both~\cite{Nakai2006}. These possibilities are discussed below.

            In an STM study on FeTe$_{0.6}$Se$_{0.4}$, the gap size away from vortices changes little from zero field up to $B_\perp$ = 6 T (Ref. \cite{Kong2019}). Within vortices, quasiparticle peaks broaden, the gap shrinks, and spectral weight appears within the gap in the form of discrete bound states \cite{Yin2015,Fu2018,Machida2018,Kong2019,Jiang2019}. The vortices themselves are arranged in an Abrikosov lattice, with a characteristic spacing that decreases with magnetic field, though the lattice can lose its periodic structure at higher fields -- around 2-3 T in Ref. \cite{Machida2018}. A planar junction measures a spatial average of the spectrum over the region of the junction, which may encompass a number of vortices which should increase with field. In NbSe$_2$, where $\Delta/E_F\ll$1, and CdGM states are densely spaced, the spatially averaged spectrum over a unit cell of the Abrikosov lattice exhibits a sub-gap spectrum which changes from a `U' shape to a `V' shape with increasing field~\cite{Nakai2006,DvirVortexState2018}. In our measurements, in contrast, a small hard gap persists to high $B$ in both orientations.
            
            In \FTS, CdGM states are located at half integer multiples of  $\Delta^2/E_F$. This should be around 650 $\mu$eV when taking $\Delta=$ 1.7 meV and $E_F=$ 4.4 meV measured by ARPES for the same stoichiometry used here~\cite{Wang2018}. Minimal energy (approximately 0.3 meV) is indeed observed by Ref.~\cite{Kong2019} in STM measurements of three vortices. Ref.~\cite{Machida2018}, on the other hand, observes discrete bound states in hundreds of vortices, finding a wide variety in energy and spacing.  
           
           Based on the above, we suggest that the spectral features observed here are the consequence of a small junction area, sensitive to up to a few vortices. According to this interpretation, the onset of the diminishing of the gap size with magnetic field at $B_\perp=$ 0.5 T could indicate that vortices begin to enter the junction at this field, suggesting a spatial scale for the junction size comparable to the inter-vortex spacing (50-100 nm). The existence of a minimal hard gap region indicates the absence of low energy in-gap states.
           
           Alternatively, an even smaller junction may be very locally sensitive to the spectrum away from the vortex center. The behaviour of the superconducting spectrum in the spatial vicinity of a vortex is described in an in depth STM study of NbSe$_2$ (Ref. \cite{kohen2006probing}). This reference reports a gradual shrinking of coherence peaks upon approaching within the scale of the London penetration depth ($\lambda$) from the vortex core. The effect is attributed to depairing due to screening currents. Upon entering the vortex core within the scale of the superconducting coherence length ($\xi$), the gap size diminishes due to the shrinking of the superconducting order parameter. This two-stage behavior qualitatively resembles our data, if we assume that a single vortex is drawing near our junction with increasing magnetic field.

        \subsection*{Bound States}
\begin{figure}[h]
            \centering
            \includegraphics[width=1\textwidth]{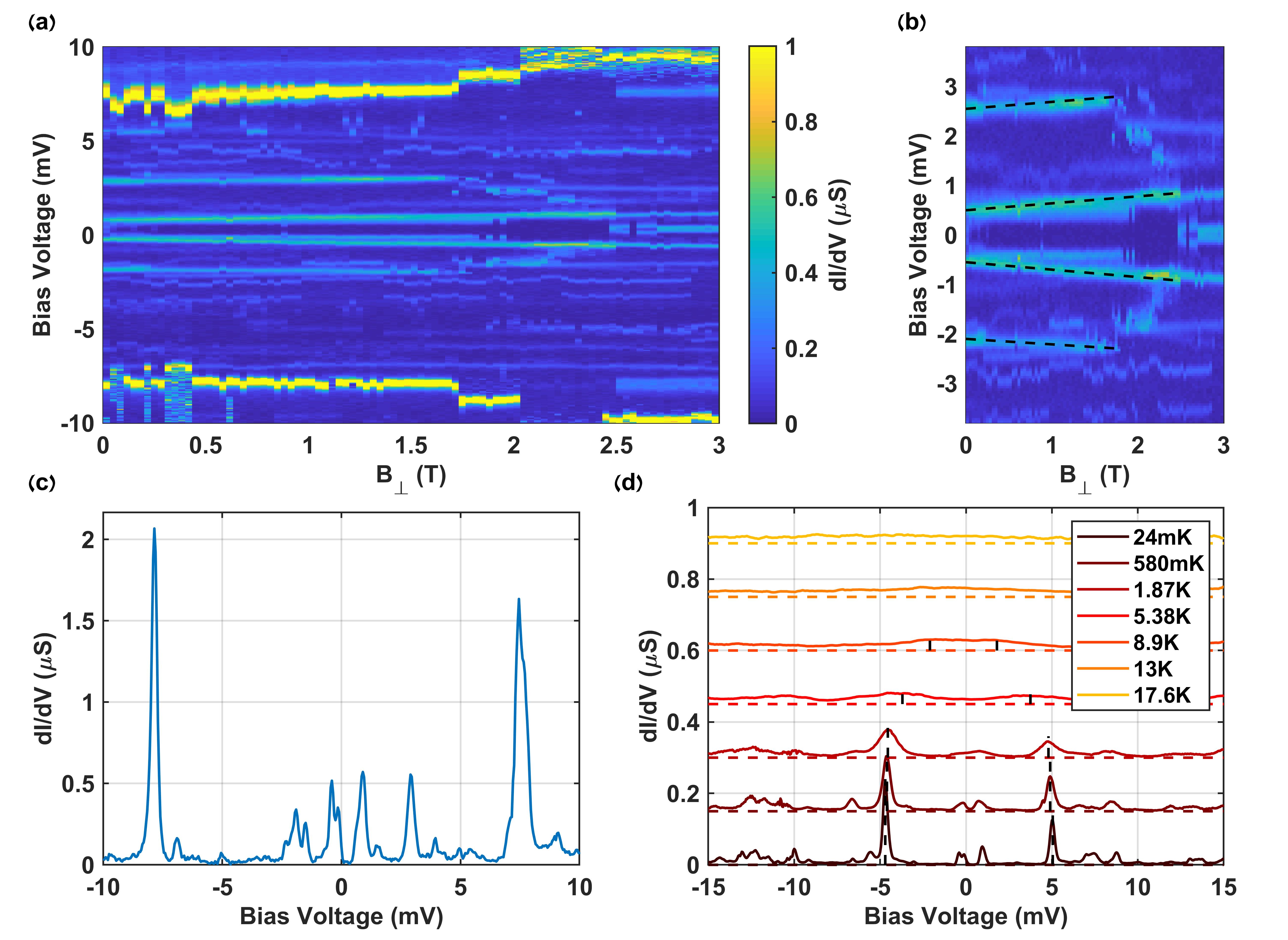}
            \caption{Spectra of top electrode Device TE: (a) Spectra at different values of $B_\perp$ (at base temperature), with no discernible superconducting gap, and displaying several bound-state-like features over and under the supposed gap. Spectral features sometimes change discontinuously with changes in the magnetic field. (b) Close-up of (a), displaying slopes of spectral features in magnetic field (symmetrized by an energy offset).
            (c) Representative line scan of the spectrum at $B_{\perp}$= 1 T (symmetrized like (b)).
            (d) Temperature dependence of the spectrum at zero field, peak locations have changed during field cycling. Smearing and closing of the two main peaks highlighted by dashed vertical lines.}
            \label{fig:states}
        \end{figure}

            We now turn to measurements taken on the top electrode tunneling device TE. Such devices are fabricated using a three-layer thick MoS$_2$ tunnel barrier placed on top of the \FTS, where top electrodes are deposited in a subsequent lithography step. The spectra of these devices (Fig. \ref{fig:states}) are dominated by pairs of sharp conductance peaks spaced nearly symmetrically about zero bias, with asymmetric conductances ranging from 0.2--2 $\mu$S. As seen in panel (c), these excitations are not restricted to sub-gap energies -- in fact, bias symmetric conductance features are found far outside the expected gap region, near the gap edge, and within the gap.
            
            We interpret these peaks as related to defect assisted tunneling into \FTS. Such defects are known to exist within the TMD tunnel barriers~\cite{Dvir2019}, as well as in the oxide layer observed by cross-section STEM measurements and within the unoxidized \FTS. There is no evidence of a superconducting gap in the spectrum, indicating that the tunneling electrode is not coupled directly to the superconductor. Several junctions on the same device exhibit qualitatively similar behavior, not presented here. 
            
            We follow the evolution of these features vs. $B_\perp$. The conductance map shows that the prominent features both inside and outside of the gap exhibit a linear evolution in energy (Fig. \ref{fig:states} panels (a,b)), with slopes ranging between $\pm$0.1-0.2 meV/T. It is noteworthy that the peaks do not split with field (this is true of the other junctions as well). The energies of the features occasionally change abruptly, while retaining energy symmetry. These sharp changes occur only in response to a change in magnetic field.
            
            The field evolution of the in-gap features appears to agree with a model where Andreev bound states (ABS) are formed by tunneling into a defect coupled to the superconductor. In this case, we interpret the observed slopes of the sub-gap linear features to originate in the Zeeman effect. Assuming a spin 1/2 defect yields a Land\'e $g$-factor of $g\approx$ 4.8-5.2, although it is possible that the defect may have a higher spin. Within the ABS model, the observation that the peaks never split with magnetic field suggests that the defect retains a doublet ground state at all fields \cite{deFran2014}. This is consistent for sub-gap features across several tunnel junctions, indicating that typically in this system, the defect charging energy is much larger than the superconducting gap. Alternatively, this could indicate that the defects are magnetic, giving rise to Yu-Shiba-Rusinov (YSR) type bound states. The existence of ABS in high magnetic fields indicates that the defect remains proximitized to the superconducting phase, and is not dominated by the vicinity of a vortex core.
            
            In panel (d), we follow the evolution of the spectrum with temperature. Increasing the temperature broadens the conductance peaks and lowers their energy, an effect highlighted by the black vertical dashed lines. The reduction of peak energy with temperature serves as an indicator that they are related to superconductivity.
            We do not currently have a convincing interpretation for the existence of symmetric features above the energies of all superconducting gaps. These features behave qualitatively similar to the sub-gap features with respect to both temperature and magnetic field, and may be related to inelastic tunneling processes, as in Ref.~\cite{Kezilebieke2018}. However, the nature of the inelastic excitations, as well as the underlying density of states, are not clear. Alternatively, resonant single particle tunneling through the defect, a process unrelated to superconductivity, may produce symmetric peaks in the unique case of equal capacitance of the defect to both source and drain. 
            
 \section*{Conclusions}
        
        We present several methods to fabricate transport and tunneling devices from \FTS. 
        Although surface oxidation was not prevented even by exfoliation in inert atmosphere, we successfully incorporate \FTS\ into devices and observe both hard gap and defect assisted tunneling. In the hard gap tunneling device, we observe the persistence of a minimal hard gap in high in-plane and out-of-plane magnetic fields. The robust nature of these gaps may be due to a small junction area, and calls for further exploration. Defect assisted tunneling into \FTS\ exhibits sub-gap resonant spectral features interpreted as ABS excitations with a doublet ground state. Less clear is the origin of the out-of-gap resonances, tentatively associated with superconductivity. These observations show that there is interest in further study of sub-gap bound states induced by proximity to an exotic superconductor. We conclude that further refinement of exfoliation and fabrication methods should enable  the investigation of iron-based superconductors by integration into vdW devices.

\section*{Methods}

    \subsection*{Pre-patterned Electrode Devices}
        The electrodes in Device BE were prepared by laser writer lithography and Ti/Au evaporation, and cleaned by plasma ashing. Flakes of MoS$_2$ (3-6 layers, identified by optical contrast) were exfoliated by the scotch tape technique and transferred to PDMS Gelpacks, selected by their optical color and transparency and stamped onto the electrodes. This technique is described in reference \cite{Castellanos-Gomez2014}. \FTS\ flakes were similarly transferred on top, so that a few electrodes are in direct contact with the flake (Ohmic contacts) while other electrodes lie underneath both the \FTS\ and the tunnel barrier, to allow for standard tunnel junction geometry. The device geometry is illustrated in Figure \ref{fig:devices_TEM}, panel (a). 
        We fabricated multiple Ohmic and tunneling contacts in this geometry. The Ohmic contacts gave resistances varying between 200-5000 $\Omega$. We note that identical devices fabricated outside of the glove box gave no Ohmic contacts, possibly due to oxidation. Tunneling contacts showed mostly trivial differential conductance, with only one contact exhibiting the tunneling behavior described in this paper.  This, too, could be due to the oxidation of the \FTS\ (see Figure \ref{fig:devices_TEM}, panel (c)).
    \subsection*{Top Electrodes}
        In Devices TR and TE, flakes of \FTS\ were exfoliated by scotch tape and transferred directly to the substrate surface. For Device TE MoS$_2$ flakes (2-7 layers, by optical contrast) were selected and stamped on top, in similar fashion to the ones in the pre-patterned electrode device. Ti/Au electrodes were fabricated on top of the heterostructure by electron beam lithography and evaporation. For the Ohmic contacts, argon ion milling was performed before evaporation to remove surface oxide. This is a crucial step for low resistance Ohmic contacts. The optimal thickness for the MoS$_2$ tunnel barrier in this method was found to be 3 layers, giving a consistent resistance of several M$\Omega$ and displaying defect assisted tunneling. Contacts on 2 layer thick flakes tended to be Ohmic while contacts on flakes thicker than 4 layers tended to have a resistance of tens of M$\Omega$ or be disconnected. We have tried fabricating devices in this geometry using WSe$_2$ as a barrier as well, and found that it usually results in Ohmic contacts. We have attempted to heat the devices after fabrication in order to clean organic residue from the interface, as works well with other materials. However, in the case of \FTS\, we have observed that heating to a temperature of 150\degree C for 3 hours produces channels in the surface, visible to an optical microscope, and a low yield of Ohmic contacts, although the material does remain superconducting in the bulk.
    \subsection*{STEM Sample Preparation and Measurement}
        Sample for the cross-section STEM measurement was prepared by using a focused ion beam, Thermo Fisher Scientific Helios NanoLab 460F1 DualBeam, to cut the a thin lamella from a sacrificial bottom electrode device. High angle annular dark field (HAADF) imaging and energy-dispersive spectroscopy (EDS) analysis was performed on Thermo Fisher Scientific Themis Z scanning transmission electron microscope (STEM).
    
\section*{Acknowledgements}
    We would like to acknowledge fruitful discussions on \FTS\ with Amit Kanigel, on YSR states with Hen Alpern, and on van der Waals tunneling into superconductors with Tom Dvir. This work was funded by a European Research Council Starting Grant (No. 637298, TUNNEL), and Israeli Science Foundation grant 1363/15. Work at Brookhaven is supported by the Office of Basic Energy Sciences, Division of Materials Sciences and Engineering, U.S. Department of Energy under Contract No. DE-SC0012704. A.Z. is grateful to the Azrieli Foundation for an Azrieli Fellowship.

\bibliographystyle{ieeetr}

\end{document}